\documentclass[11pt]{article}
\usepackage{chngcntr}
\usepackage{graphicx}

\begin{document}

\noindent\textbf{\LARGE{Age dating of an early Milky Way merger via\\ asteroseismology of the naked-eye star $\nu$\,Indi}}\\

\noindent William J. Chaplin$^{1,2,3}$, Aldo M. Serenelli$^{4,5}$, Andrea Miglio$^{1,2}$, Thierry Morel$^6$, J. Ted Mackereth$^{1,2}$, Fiorenzo Vincenzo$^{1,2,7,8}$, Hans Kjeldsen$^{2,9}$, Sarbani Basu$^{10}$, Warrick H. Ball$^{1,2}$, Amalie Stokholm$^2$,  Kuldeep Verma$^2$, Jakob R{\o}rsted Mosumgaard$^2$,   Victor Silva Aguirre$^{2}$, Anwesh Mazumdar$^{11}$, Pritesh Ranadive$^{11}$, H. M. Antia$^{12}$, Yveline Lebreton$^{13,14}$, Joel Ong$^{10}$, Thierry Appourchaux$^{15}$, Timothy R. Bedding$^{16}$, J{\o}rgen Christensen-Dalsgaard$^{2,3}$, Orlagh Creevey$^{17}$, Rafael A. Garc\'ia$^{18,19}$, Rasmus Handberg$^2$, Daniel Huber$^{20}$, Steven D. Kawaler$^{21}$, Mikkel N. Lund$^{2}$, Travis S. Metcalfe$^{22,23}$, Keivan G. Stassun$^{24,25}$, Mich{\"a}el Bazot$^{26,27}$, Paul Beck$^{28,29,30}$, Keaton J. Bell$^{31,62,23,2}$,  Maria Bergemann$^{32}$, Derek L. Buzasi$^{33}$,  Othman Benomar$^{27,34}$,  Diego Bossini$^{35}$, Lisa Bugnet$^{18,19}$, Tiago L. Campante$^{35,36}$,  Zeynep \c{C}elik Orhan$^{37}$,  Enrico Corsaro$^{38}$,  Luc\'ia Gonz\'alez-Cuesta$^{29,30}$, Guy R. Davies$^{1,2}$, Maria Pia Di Mauro$^{39}$,  Ricky Egeland$^{40}$, Yvonne P. Elsworth$^{1,2}$, Patrick Gaulme$^{23,41}$,  Hamed Ghasemi$^{42}$,  Zhao Guo$^{43,44}$,  Oliver J. Hall$^{1,2}$, Amir Hasanzadeh$^{45}$, Saskia Hekker$^{23,2}$,  Rachel Howe$^{1,2}$, Jon M. Jenkins$^{46}$, Antonio Jim\'enez$^{29,30}$, Ren\'e Kiefer$^{47}$,  James S. Kuszlewicz$^{23,2}$,  Thomas Kallinger$^{48}$, David W. Latham$^{49}$, Mia S. Lundkvist$^{2}$,  Savita Mathur$^{29,30}$,  Josefina Montalb\'an$^{1,2}$,  Benoit Mosser$^{13}$,  Andres Moya Bed\'on$^{1,2}$, Martin Bo Nielsen$^{1,2,27}$,  Sibel \"Ortel$^{37}$,  Ben M. Rendle$^{1,2}$,  George R. Ricker$^{50}$, Tha\'ise S. Rodrigues$^{51}$, Ian W. Roxburgh$^{52,1}$,  Hossein Safari$^{45}$, Mathew Schofield$^{1,2}$, Sara Seager$^{50,53,54}$, Barry Smalley$^{55}$,  Dennis Stello$^{56,16,2}$,  R\'obert Szab\'o$^{57,58}$, Jamie Tayar$^{20,63}$, Nathalie Theme{\ss}l$^{23,2}$,  Alexandra E. L. Thomas$^{1,2}$,  Roland K. Vanderspek$^{50}$, Walter E. van Rossem$^{1,2}$, Mathieu Vrard$^{35,36}$, Achim Weiss$^{59}$, Timothy R. White$^{60,2}$,  Joshua N. Winn$^{61}$, Mutlu Y\i ld\i z$^{37}$\\

\noindent $^{1}$ School of Physics and Astronomy, University of Birmingham, Birmingham B15 2TT, UK

\noindent $^{2}$ Stellar Astrophysics Centre (SAC), Department of Physics and Astronomy, Aarhus University, Ny Munkegade 120, DK-8000 Aarhus C, Denmark

\noindent $^{3}$ Kavli Institute for Theoretical Physics, University of California, Santa Barbara, CA 93106, USA

\noindent $^{4}$ Institute of Space Sciences (ICE, CSIC) Campus UAB, Carrer de Can Magrans s/n, 08193 Barcelona, Spain

\noindent $^{5}$ Institut d'Estudis Espacials de Catalunya (IEEC), C/ Gran Capit\`a, 2-4, 08034 Barcelona, Spain

\noindent $^{6}$ Space Sciences, Technologies and Astrophysics Research (STAR) Institute, Universit\'e de Li\`ege, Quartier Agora, All\'ee du 6 Ao\^ut 19c, B\^at. B5C, B4000-Li\`ege, Belgium

\noindent $^{7}$ Center for Cosmology and AstroParticle Physics, The Ohio State University, 191 West Woodruff Avenue, Columbus, OH 43210, USA

\noindent $^{8}$ Department of Physics, The Ohio State University, 191 West Woodruff Avenue, Columbus, OH 43210, USA

\noindent $^{9}$ Institute of Theoretical Physics and Astronomy, Vilnius University, Sauletekio av. 3, 10257 Vilnius, Lithuania

\noindent $^{10}$ Department of Astronomy, Yale University, P.O. Box 208101, New Haven, CT 06520-8101, USA

\noindent $^{11}$ Homi Bhabha Centre for Science Education, TIFR, V. N. Purav Marg, Mankhurd, Mumbai 400088, India

\noindent $^{12}$ Tata Institute of Fundamental Research, Mumbai, India

\noindent $^{13}$ LESIA, Observatoire de Paris, Universit\'e PSL, CNRS, Sorbonne Universit\'e, Universit\'e de Paris, 5 place Jules Janssen, 92195 Meudon, France 

\noindent $^{14}$ Univ Rennes, CNRS, IPR (Institut de Physique de Rennes) - UMR 6251, F-35000 Rennes, France

\noindent $^{15}$ Univ. Paris-Sud, Institut d'Astrophysique Spatiale, UMR 8617, CNRS, B\^atiment 121, 91405 Orsay Cedex, France

\noindent $^{16}$ Sydney Institute for Astronomy (SIfA), School of Physics, University of Sydney, NSW 2006, Australia

\noindent $^{17}$ Universit\'e C\^ote d'Azur, Observatoire de la C\^ote d'Azur, CNRS, Laboratoire Lagrange, France

\noindent $^{18}$ IRFU, CEA, Universit\'e Paris-Saclay, F-91191 Gif-sur-Yvette, France

\noindent $^{19}$ AIM, CEA, CNRS, Universit\'e Paris-Saclay, Universit\'e Paris Diderot, Sorbonne Paris Cit\'e, F-91191 Gif-sur-Yvette, France

\noindent $^{20}$ Institute for Astronomy, University of Hawai`i, 2680 Woodlawn Drive, Honolulu, HI 96822, USA

\noindent $^{21}$ Department of Physics and Astronomy, Iowa State University, Ames, IA 50011

\noindent $^{22}$ Space Science Institute, 4750 Walnut Street, Suite 205, Boulder CO 80301, USA

\noindent $^{23}$ Max-Planck-Institut f\"ur Sonnensystemforschung, Justus-von-Liebig-Weg 3, 37077 G\"ottingen, Germany

\noindent $^{24}$ Vanderbilt University, Department of Physics \& Astronomy, 6301 Stevenson Center Ln., Nashville, TN 37235, USA

\noindent $^{25}$ Vanderbilt Initiative in Data-intensive Astrophysics (VIDA), 6301 Stevenson Center Lane, Nashville, TN 37235, USA

\noindent $^{26}$ Division of Sciences, New York University Abu Dhabi, United Arab Emirates, 

\noindent $^{27}$ Center for Space Science, NYUAD Institute, New York University Abu Dhabi, PO Box 129188, Abu Dhabi, United Arab Emirates

\noindent $^{28}$ Department for Geophysics, Meteorology and Astrophysics, Institute of Physics, Karl-Franzens University of Graz, Universit\"atsplatz 5/II, 8010 Graz, Austria

\noindent $^{29}$ Instituto de Astrof\'isica de Canarias (IAC), E-38205 La Laguna,Tenerife, Spain

\noindent $^{30}$ Universidad de La Laguna (ULL), Departamento de Astrofisica, E-38206 La Laguna, Tenerife, Spain

\noindent $^{31}$ DIRAC Institute, Department of Astronomy, University of Washington, Seattle, WA 98195-1580, USA

\noindent $^{32}$ Max Planck Institute for Astronomy, 69117, Heidelberg, Germany

\noindent $^{33}$ Dept. of Chemistry and Physics, Florida Gulf Coast University, 10501 FGCU Blvd. S., Fort Myers, FL 33965 USA

\noindent $^{34}$ Solar Science Observatory, NAOJ, and Department of Astronomical Science, Sokendai (GUAS), Mitaka, Tokyo, Japan

\noindent $^{35}$ Instituto de Astrof\'isica e Ci\^encias do Espa\c{c}o, Universidade do Porto, CAUP, Rua das Estrelas, 4150-762 Porto, Portugal

\noindent $^{36}$ Departamento de F\'{\i}sica e Astronomia, Faculdade de Ci\^{e}ncias da Universidade do Porto, Rua do Campo Alegre, s/n, PT4169-007 Porto, Portugal

\noindent $^{37}$ Department of Astronomy and Space Sciences, Science Faculty, Ege University, 35100, Bornova, \.Izmir, Turkey

\noindent $^{38}$ INAF - Osservatorio Astrofisico di Catania, via S. Sofia 78, 95123, Catania, Italy

\noindent $^{39}$ INAF-IAPS, Istituto di Astrofisica e Planetologia Spaziali, Via del Fosso del Cavaliere 100, I-00133 Roma, Italy

\noindent $^{40}$ High Altitude Observatory, National Center for Atmospheric Research, 3080 Center Green, Boulder, CO, 80301, USA

\noindent $^{41}$ Department of Astronomy, New Mexico State University, P.O. Box 30001, MSC 4500, Las Cruces, NM 88003-8001, USA

\noindent $^{42}$ Department of Physics, Institute for Advanced Studies in Basic Sciences (IASBS), Zanjan 45137-66731, Iran

\noindent $^{43}$ Center for Exoplanets and Habitable Worlds, 525 Davey Laboratory, The Pennsylvania State University, University Park, PA 16802, USA

\noindent $^{44}$ Department of Astronomy \& Astrophysics, 525 Davey Laboratory, The Pennsylvania State University, University Park, PA, 16802, USA

\noindent $^{45}$ Department of Physics, University of Zanjan, P. O. Box 45195-313, Zanjan, Iran

\noindent $^{46}$ NASA Ames Research Center, Moffett Field, CA, 94035, USA

\noindent $^{47}$ Centre for Fusion, Space, and Astrophysics, Department of Physics, University of Warwick, Coventry, UK

\noindent $^{48}$ Institute of Astronomy, University of Vienna, 1180 Vienna, Austria

\noindent $^{49}$ Center for Astrophysics ${\rm \mid}$ Harvard {\rm \&} Smithsonian, 60 Garden Street, Cambridge, MA 02138, USA

\noindent $^{50}$ Department of Physics, and Kavli Institute for Astrophysics and Space Research, Massachusetts Institute of Technology, Cambridge, MA 02139, USA

\noindent $^{51}$ Osservatorio Astronomico di Padova -- INAF, Vicolo dell'Osservatorio 5, I-35122 Padova, Italy

\noindent $^{52}$ Astronomy Unit, Queen Mary University of London, Mile End Road, London, E1 4NS, UK

\noindent $^{53}$ Department of Earth, Atmospheric and Planetary Sciences, Massachusetts Institute of Technology, Cambridge, MA 02139, USA

\noindent $^{54}$ Department of Aeronautics and Astronautics, MIT, 77 Massachusetts Avenue, Cambridge, MA 02139, USA

\noindent $^{55}$ Astrophysics Group, Lennard-Jones Laboratories, Keele University, Staffordshire ST5 5BG, United Kingdom

\noindent $^{56}$ School of Physics, The University of New South Wales, Sydney NSW 2052, Australia

\noindent $^{57}$ Konkoly Observatory, CSFK, H-1121, Konkoly Thege Mikl\'os \'ut 15-17, Budapest, Hungary

\noindent $^{58}$ MTA CSFK Lend\"ulet Near-Field Cosmology Research Group

\noindent $^{59}$ Max-Planck-Institut f\"ur Astrophysik, Karl-Schwarzschild-Str. 1, D-85748 Garching, Germany

\noindent $^{60}$ Research School of Astronomy and Astrophysics, Mount Stromlo Observatory, The Australian National University, Canberra, ACT 2611, Australia

\noindent $^{61}$ Department of Astrophysical Sciences, Princeton University, Princeton, NJ 08544, USA

\noindent $^{62}$ NSF Astronomy and Astrophysics Postdoctoral Fellow and DIRAC Fellow

\noindent $^{63}$ Hubble Fellow

\vspace{10mm}

\textbf{Over the course of its history, the Milky Way has ingested multiple smaller satellite galaxies\cite{helmi18}. While these accreted stellar populations can be forensically identified as kinematically distinct structures within the Galaxy, it is difficult in general to precisely date the age at which any one merger occurred. Recent results have revealed a population of stars that were accreted via the collision of a dwarf galaxy, called \textit{Gaia}-Enceladus\cite{helmi18}, leading to a substantial pollution of the chemical and dynamical properties of the Milky Way. Here, we identify the very bright, naked-eye star $\nu$\,Indi as a probe of the age of the early in situ population of the Galaxy. We combine asteroseismic, spectroscopic, astrometric, and kinematic observations to show that this metal-poor, alpha-element-rich star was an indigenous member of the halo, and we measure its age to be $11.0 \pm 0.7$ (stat) $\pm 0.8$ (sys)$\,\rm Gyr$. The star bears hallmarks consistent with it having been kinematically heated by the \textit{Gaia}-Enceladus collision. Its age implies that the earliest the merger could have begun was 11.6 and 13.2\,Gyr ago at 68 and 95\,\% confidence, respectively. Input from computations based on hierarchical cosmological models tightens (i.e. reduces) slightly the above limits.}\\

The recently launched NASA Transiting Exoplanet Survey Satellite (TESS)\cite{ricker14} has opened the brightest stars across $\simeq 80\,\%$ of the sky\cite{stassun18} to micro-magnitude photometric studies in its two-year nominal mission. These are stars visible to the naked eye, which present huge opportunities for detailed characterization, study and follow-up.  $\nu$\,Indi (HR\,8515; HD\,211998; HIP\,110618) is a very bright (visual apparent magnitude $V=5.3$) metal-poor subgiant, which was observed by TESS during its first month of science operations. Using nearly continuous photometric data with 2-minute time sampling, we are able to measure a rich spectrum of solar-like oscillations in the star. Combining these asteroseismic data with newly analysed chemical abundances from ground-based spectroscopy, together with astrometry and kinematics from \textit{Gaia}-DR2\cite{gaia18}, show this single star as a powerful, representative tracer of old in situ stellar populations in the Galaxy. The results on $\nu$\,Indi allow us to place new constraints on the age of the in situ halo and the epoch of the \textit{Gaia}-Enceladus merger.

We re-analysed archival high-resolution spectroscopic data on $\nu$\,Indi collected by the High Accuracy Radial velocity Planet Searcher (HARPS) spectrograph\cite{mayor03} on the European Southern Observatory (ESO) 3.6-m telescope at La Silla, and by the Fiber-fed Extended Range Optical Spectrograph (FEROS)\cite{kaufer99} on the 2.2-m ESO/MPG telescope (also at La Silla). From these high-resolution spectra we measured the overall iron abundance and detailed abundances for 20 different elements, providing a comprehensive set of data on the chemistry of the star (see \textbf{Methods} for table of abundances and further details). $\nu$\,Indi exhibits enhanced levels of $\alpha$-process elements in its spectrum, i.e., elements heavier than carbon produced by nuclear reactions involving helium. The logarithmic abundance relative to iron is ${\rm [\alpha/Fe]} = +0.4$. Among Galactic disk stars, elevated ${\rm [\alpha/Fe]}$ levels are associated with old stellar populations.  $\nu$\,Indi shows an overabundance of Titanium of ${\rm [Ti/Fe]} = +0.27 \pm 0.07$, which puts it in the regime where a previous study\cite{bensby14} found ages exceeding $\approx 9.5\,\rm Gyr$ for $\alpha$-enhanced stars in the local solar neighbourhood, where $\nu$\,Indi resides.

Figure~1 shows [Mg/Fe] abundances of Milky Way stars, including $\nu$\,Indi, from the Apache Point Observatory Galaxy Evolution Experiment (APOGEE) DR-14 spectroscopic survey release\cite{Majewski17} (see \textbf{Methods} for further details).  $\nu$\,Indi's abundances place it at the upper edge of the distribution identified with the accreted \textit{Gaia}-Enceladus population\cite{helmi18} (points in red at lower [Mg/Fe]); but more in line with the in situ halo population at higher [Mg/Fe].  Were it to have been accreted, it is unlikely the star could be a member of a different accreted population, as its high [Mg/Fe] would suggest the progenitor dwarf galaxy would have had to have been at least as massive as \textit{Gaia}-Enceladus. Since the stellar debris from \emph{Gaia}-Enceladus is thought to make up a high fraction of the stellar mass of the present day halo, it seems improbable that there could exist another similar undiscovered satellite. We therefore conclude, on the basis of chemistry alone, that $\nu$\,Indi is either a member of the in-situ population, or a member of \textit{Gaia}-Enceladus. We now use kinematics to show that the former is most likely correct.


\begin{figure}
\centering
\includegraphics[width=155mm]{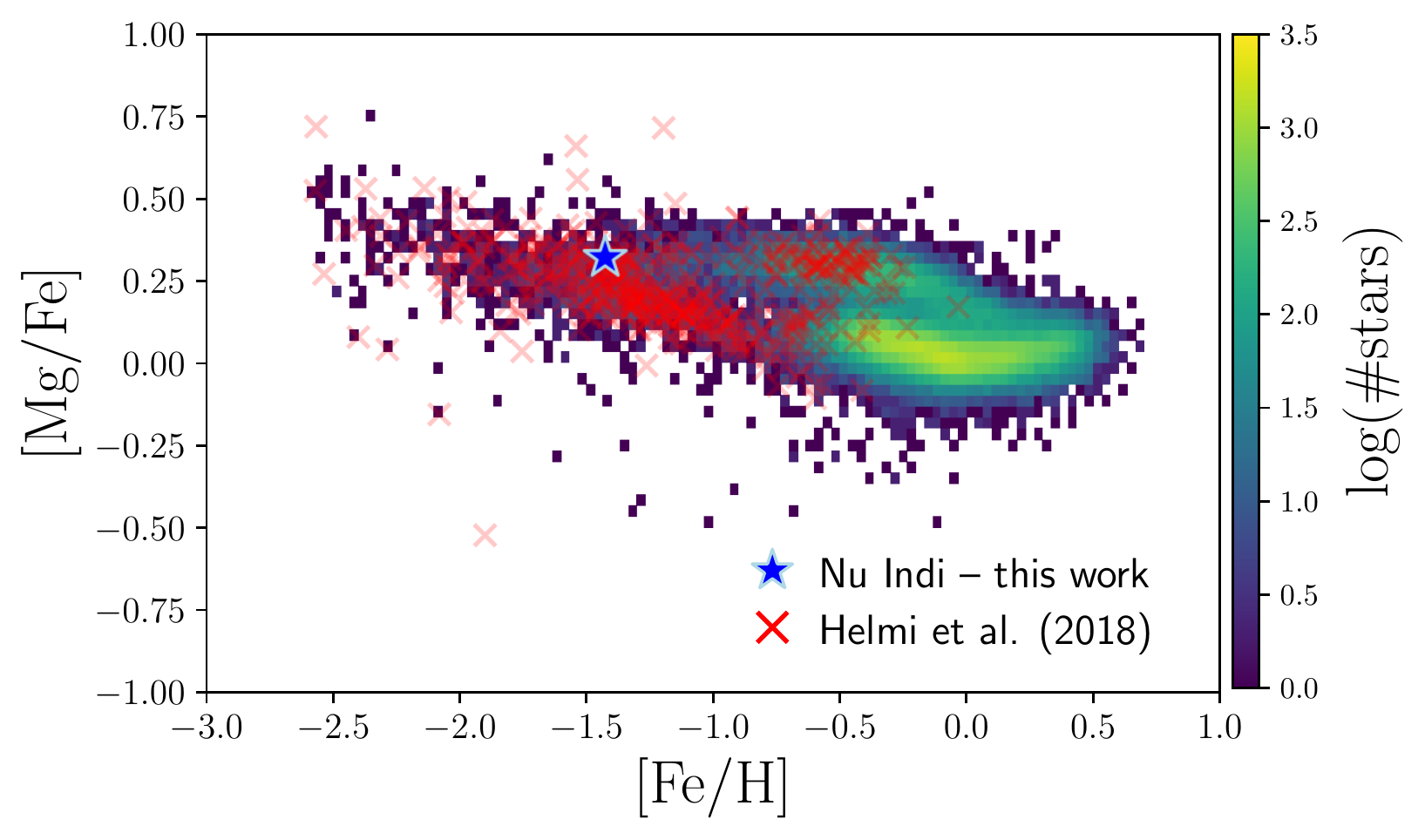}

\caption{\small [Mg/Fe] versus [Fe/H] abundances of a large sample
  of Milky Way stars, from the APOGEE DR-14 spectroscopic survey data
  release\cite{Majewski17}. Results on $\nu$\,Indi are marked by the
  blue star-shaped symbol. Points in red show the sample of stars
  identified as being part of the accreted population from
  \textit{Gaia}-Enceladus\cite{helmi18}.}

\label{fig:chem}
\end{figure}


To place $\nu$\,Indi in context among other stars with similar elemental abundances, we selected stars from APOGEE-DR14 having [Fe/H] equal (within the uncertainties) to our measured value for $\nu$\,Indi. Figure~2 shows \textit{Gaia}-DR2 velocity data for populations with low and high [Mg/Fe], which roughly divides into accreted and in situ halo stars\cite{hayes18, mackereth19}. The cross-hair marks the location of $\nu$\,Indi on both plots. The low [Mg/Fe] group includes many stars in the high-eccentricity accreted halo, which was recently determined to be dominated by the \textit{Gaia}-Enceladus accretion event. Here, the low [Mg/Fe] population shows a flat distribution (the so-called \textit{Gaia} Sausage) in the tangential versus radial velocity plane, consistent with the strong radial motion from an accreted population. In the vertical versus radial velocity plane, the distributions of the low and high [Mg/Fe] stars are remarkably similar. This suggests the in situ, higher [Mg/Fe] population, which includes $\nu$\,Indi (see below), was heated by the accreted population. We note also evidence from simulations\cite{font11,mccarthy12,tissera18} for mergers causing heating of in situ populations. 

We derived Galactic orbital parameters for $\nu$\,Indi using the positions and velocities provided by \textit{Gaia}-DR2 (see \textbf{Methods}). We performed the same orbital integrations for the populations with low and high [Mg/Fe].  Figure~3 shows a contour plot of the resulting distributions of the eccentricity, $e$, and maximum vertical excursion from the Galactic mid-plane, $z_{\rm max}$.  Low eccentricity orbits are dominated by higher [Mg/Fe] stars, and are likely part of the thick disc/in situ halo. The position of $\nu$\,Indi is marked on the contour plot; the uncertainties are too small to be visible on this scale. Our analysis of the \textit{Gaia}-DR2 data reveals that $\nu$\,Indi has a relatively eccentric orbit, with $e = 0.60 \pm 0.01$, $z_{\rm max} = 1.51 \pm 0.02\,\rm  kpc$, and a Galactic pericentric radius of $\simeq 2.5\,\rm kpc$. Given that $\nu$\,Indi lies in a region of kinematics space dominated by the higher [Mg/Fe] stars, and has an [Mg/Fe] abundance in-line with those stars, it is likely to be a member of this population, formed in situ (five times more likely, based on the data in Figures.~2 and 3).


\begin{figure}
\centering
\includegraphics[width=100mm]{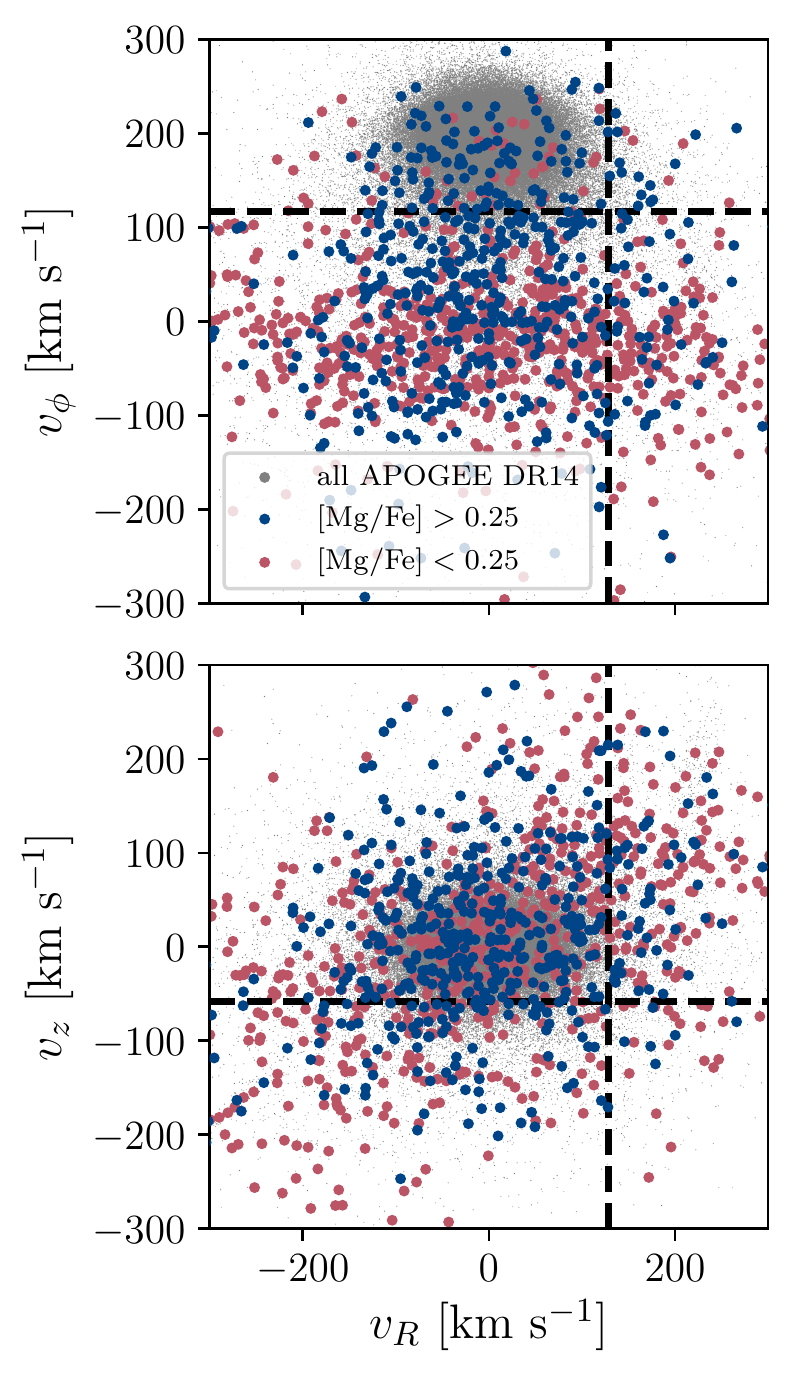}

\caption{\small Velocities of stars from APOGEE-DR14 having [Fe/H]
  lying within uncertainties of the [Fe/H] of $\nu$\,Indi. The points
  in blue show results for 637 stars with [Mg/Fe]$>+0.25$, while those
  in red are for 918 stars with [Mg/Fe]$<+0.25$. Results on the full
  APOGEE-DR14 sample are plotted in grey. Plotted, in Galacto-centric
  cylindrical coordinates and as a function of radial velocity, are
  tangential velocity (upper panel) and vertical velocity (lower
  panel). The dashed cross-hair marks the location of $\nu$\,Indi in
  these planes.}

\label{fig:kin2}
\end{figure}



\begin{figure}
\centering
\includegraphics[width=120mm]{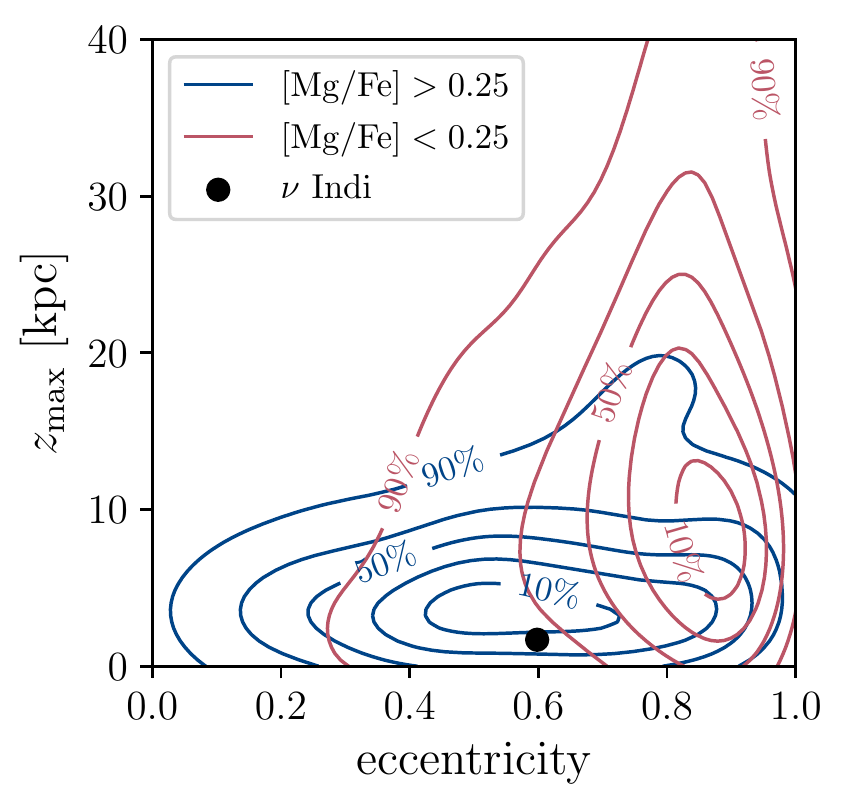}

\caption{\small Contour plot of the distribution in eccentricity, $e$,
  and maximum vertical excursion from the Galactic mid-plane, $z_{\rm
    max}$, for the same high (blue) and low (red) [Mg/Fe] samples as
  stars as Figure~\ref{fig:kin2}. The solid black symbol marks the
  location of $\nu$\,Indi. The contours are marked with the
  corresponding cumulative probabilities for each sample.}

\label{fig:kin}
\end{figure}


From our discussion above we find that $\nu$\,Indi is an in situ star whose age can provide insights on the origin of the low [Fe/H], high [Mg/Fe] population to which it belongs. The new asteroseismic data from TESS provide the means to constrain the age very precisely. $\nu$\,Indi was included on the 2-minute cadence list by the TESS Asteroseismic Science Consortium (TASC) as a prime target for asteroseismology\cite{schofield19}. It was observed for just over 27\,days in Sector\,1 of \textit{TESS} science operations. Figure~4 shows the frequency power spectrum of the calibrated lightcurve (see \textbf{Methods}).

The star shows a rich spectrum of overtones of solar-like oscillations, modes that are stochastically excited and intrinsically damped by near-surface convection\cite{chaplin13}.  The modes may be decomposed onto spherical harmonics of angular degree $l$. Overtones of radial ($l=0$), dipole ($l=1$) and quadrupole ($l=2$) modes are clearly seen. Because $\nu$\,Indi is an evolved star, its non-radial modes are not pure acoustic modes. They show so-called ``mixed'' character\cite{bedding11}, due to coupling with waves confined in cavities deep within the star for which buoyancy, as opposed to gradients of pressure, act as the restoring force.  Frequencies of mixed modes change rapidly with time as the star evolves toward the red-giant phase, and are very sensitive to the structure of the deepest lying layers providing strong diagnostic constraints on the age and structure of a star. Previous ground-based observations of precise Doppler shifts had detected solar-like oscillations in $\nu$\,Indi\cite{bedding06}, but with just a few days of data only a few oscillation modes could be identified\cite{carrier07}. With TESS, there is no ambiguity across several orders of the spectrum, and we measured precise frequencies of 18 modes spanning six overtones (see Table 1, and \textbf{Methods} for further details).


\begin{figure}
\centering
\includegraphics[width=165mm]{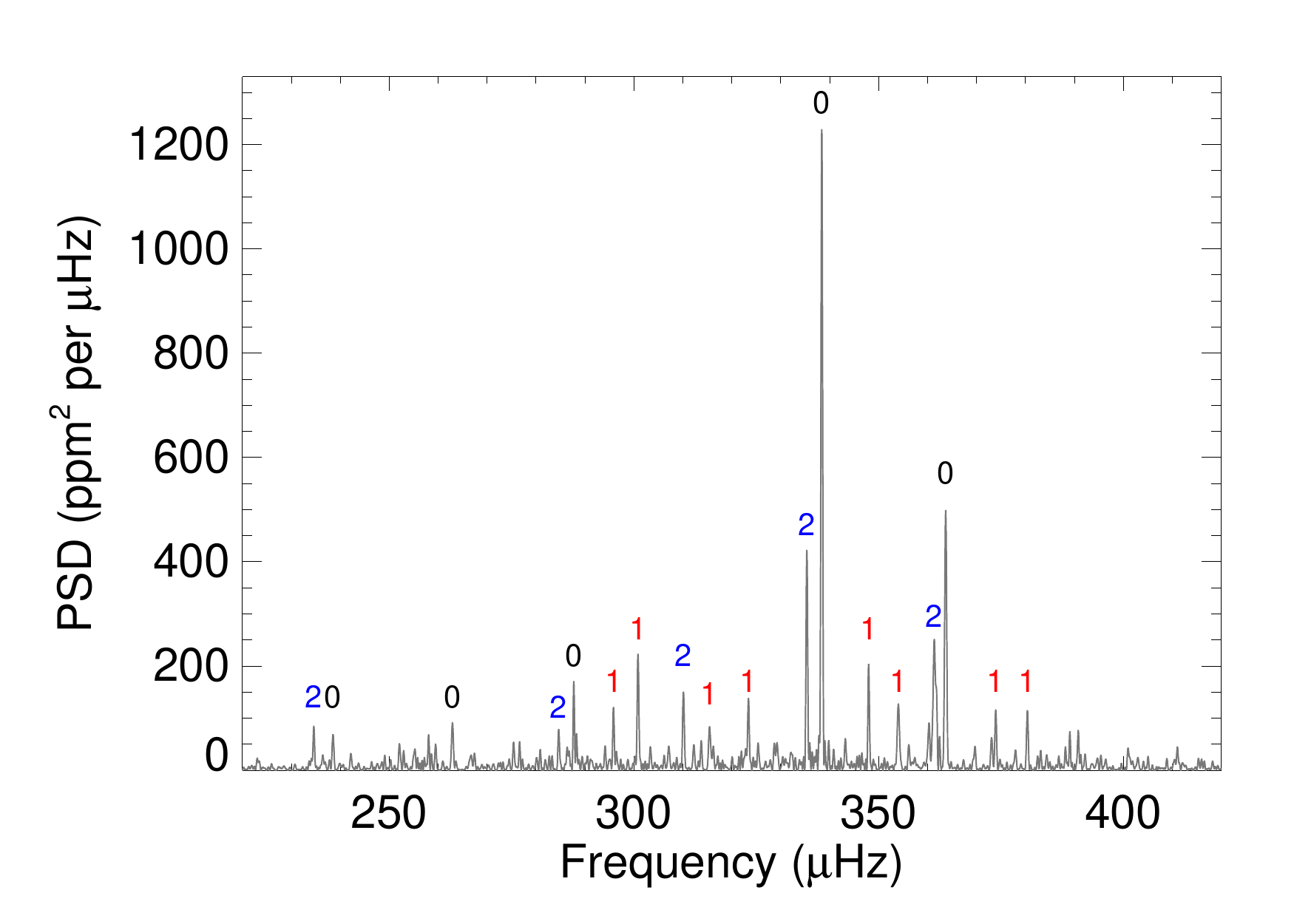}

\caption{\small Frequency-power spectrum of the TESS lightcurve of
  $\nu$\,Indi, showing a rich spectrum of solar-like
  oscillations. The ordinate is in power spectral density (PSD) units of parts per million squared per $\mu$Hz. Marked on the plot are the angular degrees, $l$, of modes whose frequencies we reported in order to model the star.}

\label{fig:osc}
\end{figure}


To constrain the mass and age of $\nu$\,Indi we used as input the measured oscillation frequencies; the spectroscopically estimated effective temperature, [Fe/H] abundance and [$\alpha$/Fe] ratio; and, as another observational constraint, the stellar luminosity given by the \textit{Gaia}-DR2 parallax and Tycho\,2\cite{hog00} $V$ and $B$-band magnitudes.  These inputs were compared, using well-developed modelling techniques\cite{serenelli17}, to intrinsic properties and predicted observables of stellar evolutionary models in evolutionary sequences sampling a dense grid in mass and composition.  We find a mass of $0.85 \pm 0.04$ (stat) $\pm 0.02$ (sys)$\,\rm M_{\odot}$ and an age of $11.0 \pm 0.7$ (stat) $\pm 0.8$ (sys)$\,\rm Gyr$. The precision achieved in mass and age is notably inferior when the asteroseismic inputs are not used.

The asteroseismic age is consistent with the claim that stars in the region of [Mg/Fe]-[Fe/H] space that includes $\nu$\,Indi were heated kinematically by the \textit{Gaia}-Enceladus merger. That episode has been estimated to have occurred between 9 and 12\,Gyr ago \cite{helmi18, vincenzo19, gallart19}. Recent results also indicate that the in situ halo was in place prior to the merger\cite{gallart19}. We may therefore use the age of $\nu$\,Indi to place a new limit on the earliest epoch at which the merger occurred (i.e., the star must have already been in place). We must take into account the uncertainty on our estimated age, and the potential duration in time of the merger itself. Numerical simulations in the literature suggest timescales for the relevant mass range of between 1 and 2\,Gyr\cite{velazquez99}. Using our posterior on the age of $\nu$\,Indi, and allowing for a spread of up to 2\,Gyr for the merger, we estimate the earliest the merger could have begun was 11.6 and 13.2\,Gyr ago at 68 and 95\,\% confidence (see \textbf{Methods} and Figures 6 and 7). The results are fairly insensitive to the merger duration (e.g., reducing the duration to 1\,Gyr reduces the 95\,\% limit by 0.3\,Gyr). Theoretical computations, based on hierarchical cosmological models (again, see \textbf{Methods}), suggest a low probability that the merger occurred before $\nu$\,Indi formed. Including this information tightens (i.e. reduces) slightly the above limits.\\

\begin{table}
\label{tab:freqs}
\centering
\caption{Measured oscillation frequencies of $\nu$\,Indi, with 1\,$\sigma$ uncertainties.}
\medskip
\begin{tabular}{ccc}
\hline
Degree, $l$& Frequency ($\mu\rm Hz$)& Uncertainty ($\mu \rm Hz$)\\
\hline
2&   234.60&  0.18\\
0&   238.52&  0.20\\
0&   262.93&  0.18\\
2&   284.62&  0.18\\
0&   287.72&  0.13\\
1&   295.81&  0.14\\
1&   300.84&  0.11\\
2&   310.10&  0.13\\
1&   315.44&  0.19\\
1&   323.41&  0.15\\
2&   335.33&  0.07\\
0&   338.38&  0.05\\
1&   347.96&  0.11\\
1&   353.98&  0.15\\
2&   361.33&  0.11\\
0&   363.70&  0.07\\
1&   373.91&  0.15\\
1&   380.39&  0.17\\
\hline
\end{tabular}
\end{table}

\newpage

\noindent \textbf{Acknowledgements} This paper includes data collected
by the TESS mission, which are publicly available from the Mikulski
Archive for Space Telescopes (MAST). Resources supporting this work
were provided by the NASA High-End Computing (HEC) Program through the
NASA Advanced Supercomputing (NAS) Division at Ames Research Center
for the production of the SPOC data products. W.J.C. acknowledges
support from the UK Science and Technology Facilities Council (STFC)
and UK Space Agency. Funding for the Stellar Astrophysics Centre is
provided by The Danish National Research Foundation (Grant agreement
no.: DNRF106). This research was partially conducted during the
Exostar19 program at the Kavli Institute for Theoretical Physics at UC
Santa Barbara, which was supported in part by the National Science
Foundation under Grant No. NSF PHY-1748958. A.M., J.T.M., F.V., and
J.M.  acknowledge support from the ERC Consolidator Grant funding
scheme (project ASTEROCHRONOMETRY, G.A. n. 772293). F.V. acknowledges
the support of a Fellowship from the Center for Cosmology and
AstroParticle Physics at The Ohio State University. W.H.B. and
M.B.N. acknowledge support from the UK Space Agency. K.J.B. is
supported by the National Science Foundation under Award
AST-1903828. M.B.N acknowledges partial support from the NYU Abu Dhabi
Center for Space Science under grant G1502. A.M.S. is partially
supported by the Spanish Government (ESP2017-82674-R) and Generalitat
de Catalunya (2017-SGR-1131). T.M. acknowledges financial support from
Belspo for contract PRODEX PLATO. H.K. acknowledges support from the
European Social Fund via the Lithuanian Science Council grant
No. 09.3.3-LMT-K-712-01-0103. S.B. acknowledges support from NSF grant
AST-1514676 and NASA grant 80NSSC19K0374. V.S.A. acknowledges support
from the Independent Research Fund Denmark (Research grant
7027-00096B). D.H. acknowledges support by the National Aeronautics
and Space Administration (80NSSC18K1585, 80NSSC19K0379) awarded
through the TESS Guest Investigator Program and by the National
Science Foundation (AST-1717000). T.S.M. acknowledges support from a
visiting fellowship at the Max Planck Institute for Solar System
Research. Computational resources were provided through XSEDE
allocation TG-AST090107. D.L.B. acknowledges support from NASA under
grant NNX16AB76G. T.L.C. acknowledges support from the European
Union's Horizon 2020 research and innovation programme under the Marie
Sk\l{}odowska-Curie grant agreement No.~792848 (PULSATION). This work
was supported by FCT/MCTES through national funds (PIDDAC) by means of
grant UID/FIS/04434/2019. K.J.B., S.H., J.S.K. and N.T. are supported
by the European Research Council under the European Community's
Seventh Framework Programme (FP7/2007-2013)/ERC grant agreement no
338251 (StellarAges). E.C. is funded by the European Union's Horizon
2020 research and innovation program under the Marie Sklodowska-Curie
grant agreement No. 664931. L.G.C. acknowledges support from the
MINECO FPI-SO doctoral research project SEV-2015-0548-17-2 and
predoctoral contract BES-2017-082610. P.G. is supported by the German
space agency (Deutsches Zentrum f\"ur Luft- und Raumfahrt) under PLATO
data grant 50OO1501. R.K. acknowledges support from the UK Science and
Technology Facilities Council (STFC), under consolidated grant
ST/L000733/1. M.S.L. is supported by the Carlsberg Foundation (Grant
agreement no.: CF17-076)". Z.C.O., S.O. and M.Y. acknowledge support
from the Scientific and Technological Research Council of Turkey
(T\"UB\.ITAK:118F352). S.M. acknowledges support from the Spanish
ministry through the Ramon y Cajal fellowship number RYC-2015-17697.
T.S.R. acknowledges financial support from Premiale 2015 MITiC (PI
B. Garilli). R. Sz. acknowledges the support from NKFIH grant project
No. K-115709, and the Lend\"ulet program of the Hungarian Academy of
Science, project No. 2018-7/2019.  J.T. acknowledges support was
provided by NASA through the NASA Hubble Fellowship grant No. 51424
awarded by the Space Telescope Science Institute, which is operated by
the Association of Universities for Research in Astronomy, Inc., for
NASA, under contract NAS5-26555. This work was supported by FEDER
through COMPETE2020 (POCI-01-0145-FEDER-030389. A.M.B. acknowledges
funding from the European Union’s Horizon 2020 research and innovation
program under the Marie SklodowskaCurie grant agreement No 749962
(project THOT). A.M. and P.R. acknowledge the support of the Govt. Of
India, Department of Atomic Energy, under Project
No. 12-R\&D-TFR-6.04-0600.\\

\noindent \textbf{Author contributions} W.J.C. led the project, with
help from A.M.S., A.M., S.B. and W.H.B. W.J.C., H.K., W.H.B., H.M.A.,
T.R.B., R.A.G., D.H., K.J.B., D.L.B., O.B., L.B., T.L.C., E.C.,
L.G.-C., G.R.D., Y.P.E., P.G., H.G., O.J.H., A.H., S.H., R.H., A.J.,
R.K., J.S.K., T.K., M.S.L., S.M., B.M., A.M.B., M.B.N., I.W.R., H.S.,
R.S., N.T., A.E.L.T., M.V. and T.M.W. worked on extracting mode
parameters from the TESS data. R.H. and M.N.L. oversaw production of
the TESS lightcurves for the asteroseismic analysis. A.M.S., A.M.,
S.B., W.H.B., A.S., K.V., J.R.M., V.S.A., A.M., P.R., Y.B., J.O.,
P.B., M.B., K.J.B., D.B., Z.C.O., M.P.D.M., Z.G., S.H., J.M., S.O.,
B.M.R., T.S.R., D.S., J.T., W.E.v.R., A.W. and M.Y. worked on
modelling $\nu$\,Indi. T.M. performed the spectroscopic analysis of
the archival HARPS and FEROS data on $\nu$\,Indi. M.B. assessed the
impact of NLTE on the spectroscopic analysis. R.E. performed the
chromospheric activity analysis of $\nu$\,Indi. J.T.M. performed the
kinematics analysis and comparison of the chemistry of $\nu$\,Indi
with samples of Milky Way stars, and F.V. computed the theoretical
prior based on hierarchical cosmological models of structure
formation. D.H., K.G.S. and B.S. provided estimates of the luminosity
of $\nu$\,Indi. J.C.-D., H.K., W.J.C., T.R.B., S.D.K., S.B. are key
architects of TASC (members of its board), whilst G.R.R., J.M.J.,
D.W.L., R.K.V., J.N.W. are key architects of the TESS Mission. W.J.C.,
D.H., T.A., A.M.S., O.C., R.A.G. and T.S.M. oversaw the TASC working
groups on solar-like oscillators, and with M.S. and T.L.C., oversaw
the selection of short-cadence targets for asteroseismic studies of
solar-like oscillators with TESS, which included ensuring $\nu$\,Indi
was included on the list (and hence received the TESS short-cadence
data needed to make this study possible). All authors have contributed
to the interpretation of the data and the results, and discussion and
giving comments on the paper.\\

\noindent \textbf{Competing Interests} The authors declare that they have no competing financial interests.\\

\noindent \textbf{Correspondence} Correspondence and requests for materials should be addressed to W.J.C.. (email: w.j.chaplin@bham.ac.uk).

\newpage


\noindent\textbf{\LARGE{Methods}}\\

\noindent\textbf{Spectroscopic analysis}\\
The results of our detailed spectroscopic analysis are presented in Table~\ref{tab:spec}

We base the analysis primarily on the average of six HARPS spectra obtained in 2007 December, retrieved from the instrument archives. They have a resolving power, $R$, of 115\,000 and cover the spectral domain from 379 to 691\,nm (with a gap between 530.4 and 533.8\,nm). The signal-to-noise ratio, S/N, at 550\,nm lies in the range 177 to 281. We carried out a differential, line-by-line analysis relative to the Sun. The high-quality (S/N$\sim$470) solar \textit{HARPS} spectrum was taken from the online library of \textit{Gaia} FGK benchmarks\cite{blanco14m}. It is a solar reflected spectrum from asteroids with a similar resolution to that of the spectra for $\nu$\,Indi. For oxygen we made use  of the O\textsc{I} triplet at $\sim$777.4\,nm. Because this range is not covered by the HARPS spectra, we used the spectrum available in the FEROS archives ($R$ $\sim$ 47\,000 and a mean S/N of 340). For the Sun, numerous asteroid spectra were considered. All the spectra were normalised to the continuum by fitting low-order cubic spline or Legendre polynomials to the line-free regions using standard tasks implemented in the IRAF software\cite{tody86m}.
  
The stellar parameters and abundances of 20 elements were determined self-consistently from the spectra, plane-parallel MARCS model atmospheres\cite{gustafsson08m}, and the 2017 version of the line-analysis software MOOG. We used a line list\cite{chen00m} augmented\cite{melendez14m, reddy03m} for C\,I, Sc\,II, Mn\,I, Co\,I, Cu\,I, Zn\,I, Y\,II, and Zr\,II. Equivalent widths (EW) were measured manually assuming Gaussian profiles. Only lines above 480.0\,nm were considered because of strong line crowding in the blue that leads to an uncertain placement of the  continuum. With the exception of Mg\,I $\lambda$571.1, lines with relative width RW = log(EW/$\lambda$) $>$ --4.8 were discarded. Hyperfine structure (HFS) and isotopic splitting were taken into account for Sc, V, Mn, Co, and Cu using atomic data from the Kurucz database with an assumed Cu isotopic ratio\cite{asplund09m}. The {\tt blends} driver in MOOG was employed for the analysis. The corrections are very small for $\nu$\,Indi, but can be substantial for the Sun. The determination of the Li and O abundances from Li\,I $\lambda$670.8 and [O\,I] $\lambda$630.0 relied on a spectral synthesis\cite{morel14m}, taking the macroturbulent and projected rotational velocities of $\nu$\,Indi into account\cite{bruntt10m}.

The four model parameters --- effective temperature $T_\mathrm{eff}$, surface gravity $\log g$, metallicity [Fe/H] and microturbulence parameter $\xi$ --- were modified iteratively  until the excitation and ionization balance of iron was fulfilled and the Fe\,I abundances exhibited no trend with RW. The abundances of iron and the $\alpha$ elements were also required to be consistent with the values adopted for the model atmosphere. For the solar analysis, $T_\mathrm{eff}$ and $\log\,g$ were held fixed at 5777\,K and 4.44\,dex, respectively, whereas the microturbulence, $\xi$, was left as a free parameter (we obtained $\xi_{\odot}$ = 0.97 km s$^{-1}$). We also performed the analysis with the surface gravity of $\nu$\,Indi fixed to the asteroseismic value of $\log g = 3.46$\,dex in order  to increase both the accuracy and precision of the spectroscopic results. For this constrained analysis, we adjusted $T_\mathrm{eff}$ to satisfy iron ionization equilibrium. 

The uncertainties in the stellar parameters and abundances were computed following well-established procedures\cite{morel18m}. In particular, the analysis was repeated using Kurucz atmosphere models and the differences incorporated in the error budget. However, the deviations with respect to the default values (Kurucz minus MARCS) appear to be small: $\Delta T_\mathrm{eff}$ = --15 K, $\Delta \log g$ = --0.01, and abundance ratios deviating by less than 0.01\,dex.

We also computed corrections to the abundances for non-LTE (NLTE) effects, with those corrections defined as the difference in abundance
required to fit a line profile using NLTE or LTE models,
respectively. The NLTE corrections were estimated for most of the
spectral lines in the LTE analysis using the interactive online
tool at \texttt{nlte.mpia.de}. Corrections for $\nu$\,Indi were computed
using a MARCS model atmosphere.  We also computed corrections for
the Sun, but using a more appropriate MAFAGS-OS model, and subtracted
the solar corrections from the corrections for $\nu$\,Indi in order
to compensate for the LTE minus NLTE differences in the reference
regime. Note the difference between MARCS and MAFAGS
is negligible for main-sequence stars stars\cite{bergemann12m}.

We used the online tool to compute corrections for O, Mg, Si, Ca,
and Cr. The data used are based on the NLTE model
\textit{atoms}\cite{bergemann12m, bergemann10am, bergemann10bm,
bergemann13m, bergemann17m}. NLTE corrections for the lines of Mn
were computed separately\cite{bergemann08m, bergemann19m}, as these atoms are not yet a part of the
publicly released grid that is coupled to the online tool. For
several elements, no NLTE data are available
in the literature.

We found corrections that are typically within the quoted
abundance uncertainties -- for example, the correction to the overall
Iron abundance [Fe/H] was 0.07 -- which do not have a substantial impact on the estimated fundamental properties of the star.

The above analyses yielded an estimated effective temperature of $T_{\rm eff} = 5320 \pm 24\,\rm K$ from the asteroseismically constrained analysis and $T_{\rm eff} = 5275 \pm 45\,\rm K$ from the unconstrained analysis; and a NLTE-corrected metallicity of [Fe/H]$=-1.43 \pm 0.06$ from the constrained analysis, and [Fe/H]$=-1.46 \pm 0.07$ from the unconstrained analysis. Detailed chemical abundances are listed in Table~2. The values in brackets give the number of features each abundance is based on. For iron, the number of Fe\,I and Fe\,II lines is given. The final iron abundance is the unweighted average of the Fe\,I and Fe\,II values. For oxygen, we adopt the value given by [O\,I] $\lambda$\,630 because it is largely insensitive to non-LTE and 3D effects.\\

We also analyzed the chromospheric activity of $\nu$ Indi using 116 archival Ca HK spectra from the SMARTS Southern HK program, 
obtained 2007--2012.  The median $S$-index calibrated to the Mount
Wilson scale is 0.138, which is converted to the bolometric-relative 
HK flux $\log(R'_{\rm HK}) = -5.16$ using an empirical relation\cite{noyes84m} and the color index $B-V = 0.65$. This is in good agreement with other results in the literature \cite{henry96m}. Chromospheric
activity is a well-known proxy for age, and this low value is consistent with a very old star\cite{wright2004m}.  The empirical age-activity relationship\cite{mamajek08m} is calibrated to a low activity limit of $\log(R'_{\rm HK}) = -5.10$, corresponding to lower limit age of 8.4\,Gyr with an estimated uncertainty of 60\%, consistent with the result from our asteroseismic analysis.\\

\noindent\textbf{APOGEE-DR14 and \textit{Gaia}-DR2 analysis}\\
To construct Figure~1 of the main paper, we used abundances from the fourteenth data release (DR-14) of the SDSS\,IV-APOGEE survey, which obtained high resolution ($R \simeq 20,000$), high signal-to-noise ratio (${\rm SNR} \simeq 100$\,per pixel) spectra in the near infrared H-band. We take the calibrated [Fe/H] and [Mg/Fe] abundances directly from the APOGEE DR-14 catalogue, selecting only stars which form part of the main survey (i.e. part of the "statistical sample"). We also performed a cross match between this catalogue and the stars identified\cite{helmi18m} as being part of the \textit{Gaia}-Enceladus population on the basis of their angular momenta (as measured using \textit{Gaia}-DR2 data); as such, this population is likely contaminated by thick disk stars, which have considerably higher [Fe/H] and [Mg/Fe] than the true \textit{Gaia}-Enceladus populations.

For the kinematics analysis (Figures 2 and 3 of the main paper), we used the six-dimensional information (positions and velocities) provided by \textit{Gaia}-DR2 to derive Galactic orbital parameters for $\nu$\,Indi, as well as stars from APOGEE-DR14 having [Fe/H] equal (within the uncertainties) to our measured value for $\nu$\,Indi. APOGEE stars were targeted\cite{zasowski13m, zasowski17m} based on their ($J-K$) color and $H$-band magnitude alone, and so the selection does not result in any substantial kinematic biases to the data.  More than 90\,\% of the APOGEE stars we selected have a \emph{Gaia}-DR2 proper motion.

By reconstructing and taking samples from the covariance matrix of the astrometric parameters, we performed orbital integrations from 1000 realisations of the initial phase-space coordinates of the star. We used the python package \texttt{galpy}\cite{bovy15m}, adopting a Milky-Way-like potential (having verified that reasonable changes to the potential did not affect the conclusions drawn from our results). To convert between the observed astrometric parameters (positions, parallaxes, proper motions, and radial velocities) and Galactocentric positions and velocities we adopted the Galactocentric distance of the GRAVITY collaboration\cite{gravity18m} of 8.127\, kpc, the height $z_0=0.02\,\rm kpc$ of the Sun above the midplane of the Galaxy\cite{bennett19m}, and a solar velocity from a recent re-assessment of the stellar kinematics of the solar neighbourhood\cite{schoenrich10m}.\\

\noindent\textbf{Asteroseismic analysis}\\
The TESS target pixel file data for $\nu$\,Indi were produced by the TESS Science Operations Center (SPOC)\cite{jenkins16m}, and are available at the Mikulski Archive for Space Telescopes (MAST)\\ (http://archive.stsci.edu/). The lightcurve we analysed was extracted from target pixel files by the TESS Asteroseismic Science Operations Centre (TASOC) pipeline\cite{lund17m}. A rich spectrum of overtones of radial- and non-radial solar-like oscillations is clearly detectable (see Figure~4 of the main paper). Even though the modes are intrinsically damped, the lifetimes are longer than the 27-day length of the TESS data. The modes may as such be treated as being coherent on the timescale of the lightcurve, and we extracted their frequencies using a well-tested weighted sine-wave fitting analysis\cite{kjeldsen05m, bedding07m}, which allowed for the varying quality of the TESS photometry over the period of observation. Approaches based on fitting Lorentzian-like models to the resonant peaks\cite{benomar09m, gaulme09m, mosser12m, corsaro14m, corsaro15m, vrard15m, nielsen17m, roxburgh17m, montellano18m, benomar18m, kallinger18m} gave very similar results. Corrections to the frequencies to allow for the line-of-sight velocity of the star\cite{davies2014m} are very small, and do not change the inferred stellar properties. The list of frequencies, together with equivalent 1\,$\sigma$ uncertainties, is presented in Table~1 of the main paper.

The oscillation frequencies were used as input to the stellar modelling, along with spectroscopically derived effective temperature $T_{\rm eff}$, metallicity [Fe/H], and $\alpha$-enhancement, [$\alpha$/Fe], all from the asteroseismically constrained analysis, and an estimate of the stellar luminosity $L = 6.00 \pm 0.35\,\rm L_{\odot}$, using the \textit{Gaia}-DR2 parallax and Tycho\,2 $V$ and $B$-band magnitudes\cite{hog00m}, and a bolometric correction appropriate to the $\alpha$-enhanced composition\cite{casagrande18m} (and assuming negligible extinction). We note that a Spectral Energy Distribution (SED) fit\cite{stassun16m} gave similar constraints on luminosity.

Prior to use in the modelling we inflated the uncertainties on $T_{\rm eff}$ and [Fe/H] to account for systematic differences between spectroscopic methods by adding, respectively, 59\,K and 0.062 in quadrature to the formal uncertainties\cite{torres12m}, yielding final values of $T_{\rm eff} = 5320 \pm 64\,\rm K$ and [Fe/H]$=-1.43 \pm 0.09$. 

$\nu$\,Indi is a metal-poor star showing noticeable $\alpha$ enhancement, which affects the mapping of [Fe/H] to the metal-to-hydrogen abundance ratio $Z/X$. Some modellers used grids of stellar evolutionary models that did not include the requisite enrichment, and under such circumstances a correction must be applied to the raw [Fe/H] to allow it to be used in modelling using those grids. Here, the correction needed\cite{salaris93m} is $+0.25$. This gave a corrected metallicity of [Fe/H]$=-1.18 \pm 0.11$, where the error bar was inflated further to account for uncertainty in the correction.

Various codes\cite{rendle19m, ong18m, serenelli17m, silva17m, mosumgaard18m, ball17m, lebreton14m, yildiz16m} were used to model the star and to explore its fundamental stellar properties. $\nu$\,Indi is in a rapid stage of stellar evolution, and we found it was imperative that the codes interrogated model grids sampled at a fine resolution in mass and metallicity in order to obtain a good match of predicted observables of the best-fitting model to the actual observables. Our best-fitting estimates are $0.85 \pm 0.04$ (stat) $\pm 0.02$ (sys)$\,\rm M_{\odot}$ and an age of $11.0 \pm 0.7$ (stat) $\pm 0.8$ (sys)$\,\rm Gyr$. The central values and statistical uncertainties were provided by one of the codes\cite{serenelli17m}, which returned the best match to the input data. The systematic uncertainties reflect the scatter between different results. In all cases, the errors correspond to a 68\,\% confidence level.

Figure~5 is an \'echelle diagram showing the match between the observed frequencies (in grey) and the best-fitting model frequencies (coloured symbols). 

We also tested the impact of removing the asteroseismic frequencies from the modelling. This inflated the fractional uncertainty on the mass (stat) from $\simeq 5\,\%$ to $\simeq 8\,\%$, and the fractional uncertainty on age from less than 10\,\% to more than 30\,\%.\\

\noindent\textbf{\textit{Gaia}-Enceladus epoch analysis}\\
Our estimated age for $\nu$\,Indi was used to place a new limit on the earliest epoch at which the \textit{Gaia}-Enceladus merger occurred.  This took into account the uncertainty on the estimated age, and the potential duration in time of the merger itself. Figures 6 and 7 capture these results, as we explain below.

To place constraints on the duration of the merger, we estimated the dynamical friction timescale for the orbit of \textit{Gaia}-Enceladus to decay due to the drag force exerted on it by the diffuse dark matter halo of the Milky Way. We adopted a widely-used formulation\cite{lacey93m}, assumed that at the epoch of the merger the mass ratio between \textit{Gaia}-Enceladus and our Galaxy was one-quarter\cite{helmi18m}, and that the orbit of \textit{Gaia}-Enceladus was strongly radialised\cite{belokurov18m}. This procedure gave a merger timescale of less than or around 1\,Gyr. Numerical simulations in the literature suggest timescales for the relevant mass range that are between 1 and 2\,Gyr\cite{velazquez99m}. Here, we adopt the largest value of 2\,Gyr.

To estimate the limit on the epoch of the merger we started from the probability distribution on the age of $\nu$\,Indi but considered as the cumulative probability distribution function that expresses the probability of the existence of the star at any given epoch (plotted as a dashed line in Figures~6 and 7). The probability tends to unity at epochs more recent than the central age estimate, and to zero at epochs earlier than the central age estimate. (Note we combined the statistical and systematic errors in quadrature, so that the distribution is described by a mean of 11\,Gyr and a standard deviation of 1.1\,Gyr.) If the merger was instantaneous, the above distribution function would give us the sought-for limit on the earliest possible epoch. But it is not, and so we used a Gaussian distribution to describe the merger, having a FWHM of up to 2\,Gyr. We may consider this function as describing  the probability of interaction of the merger with $\nu$\,Indi. When convolved with the cumulative age probability distribution of the star, we obtain the cumulative probability for the merger (solid black line in Figures~6 and 7), and limits on the earliest epoch of merger of $11.6\,\rm Gyr$ ago at 68\,\% confidence, and 13.2\,Gyr ago at 95\,\% confidence.

We then folded in a theoretical prior on the probability of occurrence of the merger at different epochs, based on hierarchical cosmological models of structure formation. We estimated a cumulative prior probability using the Press-Schechter formalism\cite{lacey93m,mo10m}, as the conditional cumulative probability $\mathcal{P}( t < t_{\rm{merg}} ) = \mathcal{P}( M_{\rm{MW}}, t < t_{\rm {merg}} | M_{\rm{Enc}}, t_{\rm{Enc}} )$ that the Enceladus dark matter halo (of mass $M_{\rm{Enc}}$) formed at the time $t_{\rm{Enc}}$ and was later incorporated into the larger Milky Way dark matter halo (of mass $M_{\rm{MW}}$) already in place at the time of the merger $t = t_{\rm{merg}}$, which is the independent variable in our computation. We assumed values for the virial mass of the \textit{Gaia}-Enceladus dark matter halo between a lower limit of $M_{\rm Enc} = 1 \times 10^{10}\,\rm M_{\odot}$\cite{belokurov18m} and $1 \times 10^{11}\,\rm M_{\odot}$\cite{belokurov18m,myeong19m}, formed at the cosmic time $t_{\rm Enc} = 1.5\,\rm Gyr$ which corresponds to the observed median age of \textit{Gaia}-Enceladus stars\cite{gallart19m}. Finally we assumed that at the epoch of merger the Milky Way dark matter halo had a Virial mass $M_{\rm MW} = 4 \times 10^{11}\,\rm M_{\odot}$, which has been derived at redshift $z=2$ from the predicted cosmological halo mass accretion history of a Milky Way like galaxy\cite{correa15am,correa15bm,correa15cm}.   

Priors are plotted as a dot-dashed line for $M_{\rm Enc} = 1 \times 10^{10}\,\rm M_{\odot}$ in Figure~6, and $1 \times 10^{11}\,\rm M_{\odot}$ in Figure 7. Both suggest there was a low probability of the merger occurring prior to the formation of $\nu$\,Indi. Including the prior, we obtain the cumulative probabilities for the merger shown by the red lines in both figures, which tighten the limiting epoch (at 95\,\% confidence) to 11.7\,Gyr for $M_{\rm Enc} = 1 \times 10^{10}\,\rm M_{\odot}$ (Figure 6), and 12.4\,Gyr for $M_{\rm Enc} = 1 \times 10^{11}\,\rm M_{\odot}$ (Figure 7). We also tested the impact of varying $t_{\rm Enc}$ by a $\pm 1\,\rm Gyr$, and using a Milky Way mass up to $10^{12}\,\rm M_{\odot}$. These variations gave changes of up to $\simeq 0.5\,\rm Gyr$ in the inferred limit on the merger epoch; but overall the tendency is to tighten the limit obtained without the prior.

\begin{table}
\centering
\caption{Spectroscopically derived abundances and 1\,$\sigma$ uncertainties, without (unconstrained) and with (constrained) an asteroseismic constraint on log\,{\it g}. Values in brackets give the number of features each abundance is based on. For iron, the number of Fe I and Fe II lines is given. The final iron abundance is the unweighted average of the Fe I and Fe II based values. Abundances corrected for NLTE effects are marked by an asterisk.}
\medskip
\begin{tabular}{lcc}
\hline
Element&   Unconstrained&     Constrained\\
       & abundance      & abundance      \\
\hline
$[$Fe/H$]$*                               & --1.46$\pm$0.07 (58,5) & --1.43$\pm$0.06 (58,5)\\
$[$Li/H$]$                               & --0.01$\pm$0.09 (1)                     &  +0.04$\pm$0.07 (1)\\
$[$C/Fe$]$                               &  +0.33$\pm$0.09 (1)                     &  +0.31$\pm$0.08 (1)\\
$[$O/Fe$]$ (O\,I)*                  &  +0.60$\pm$0.10 (2)                     &  +0.56$\pm$0.09 (2)\\
$[$O/Fe$]$ ([O\,I])*                &  +0.41$\pm$0.09 (1)                     &  +0.45$\pm$0.08 (1)\\
$[$Na/Fe$]$                              & --0.20$\pm$0.10 (2)                     & --0.21$\pm$0.10 (2)\\
$[$Mg/Fe$]$*                              &  +0.34$\pm$0.08 (1)                     &  +0.32$\pm$0.08 (1)\\
$[$Si/Fe$]$*                              &  +0.18$\pm$0.06 (7)                     &  +0.17$\pm$0.06 (7)\\
$[$Ca/Fe$]$*                              &  +0.41$\pm$0.07 (6)                     &  +0.40$\pm$0.06 (6)\\
$[$Sc/Fe$]$                              &  +0.00$\pm$0.06 (2)    &  +0.02$\pm$0.06 (2)\\
$[$Ti/Fe$]$                              &  +0.27$\pm$0.07 (4)                     &  +0.27$\pm$0.07 (4)\\
$[$V/Fe$]$                               &  +0.00$\pm$0.12 (3)    &  +0.02$\pm$0.11 (3)\\
$[$Cr/Fe$]$*                              & --0.13$\pm$0.08 (1)                     & --0.14$\pm$0.08 (1)\\
$[$Mn/Fe$]$*                              & --0.23$\pm$0.08 (3)    & --0.23$\pm$0.07 (3)\\
$[$Co/Fe$]$                              &  +0.18$\pm$0.10 (3)    &  +0.19$\pm$0.09 (3)\\
$[$Ni/Fe$]$                              & --0.08$\pm$0.07 (13)                    & --0.08$\pm$0.07 (13)\\
$[$Cu/Fe$]$                              & --0.38$\pm$0.08 (1)    & --0.39$\pm$0.08 (1)\\
$[$Zn/Fe$]$                              &  +0.16$\pm$0.09 (1)                     &  +0.15$\pm$0.09 (1)\\
$[$Y/Fe$]$                               &  +0.08$\pm$0.07 (3)                     &  +0.10$\pm$0.07 (3)\\
$[$Zr/Fe$]$                              &  +0.38$\pm$0.08 (1)                     &  +0.40$\pm$0.08 (1)\\
$[$Ba/Fe$]$                              & --0.02$\pm$0.13 (2)                     &  +0.00$\pm$0.13 (2)\\
\hline
\end{tabular}
\label{tab:spec}
\end{table}

\newpage


\begin{figure}
\centering
\includegraphics[width=165mm]{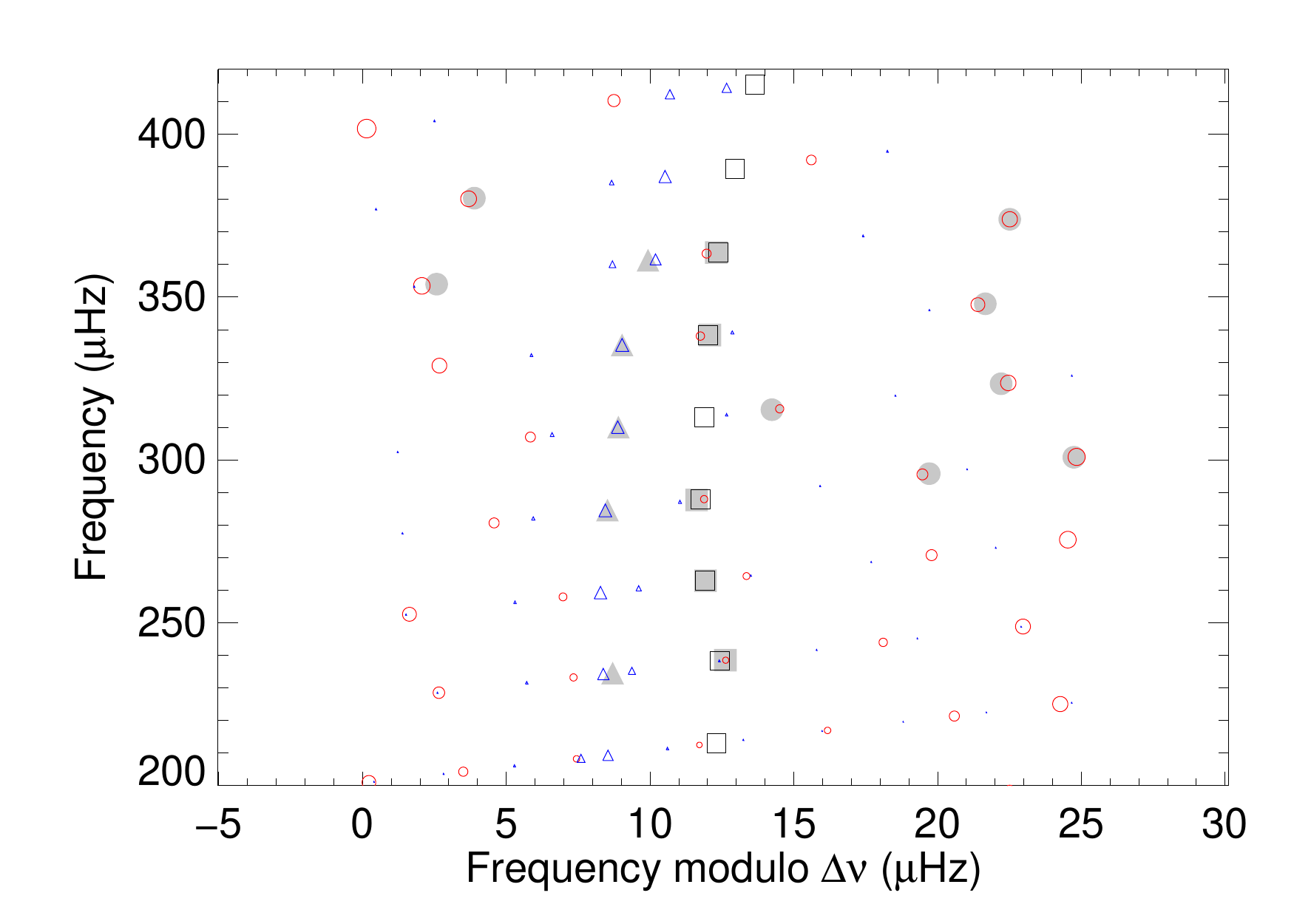}

\caption{\small An \'echelle diagram showing the observed frequencies
  (in grey) and the best-fitting model frequencies (coloured
  symbols). The diagram was made by dividing the spectrum into
  segments of length equal to the average frequency separation
  $\Delta\nu$ between consecutive overtones, which were then stacked
  in ascending order, so one plots $\nu$ versus ($\nu$ mod
  $\Delta\nu$). The $l=0$ (radial) modes are plotted with square
  symbols, the $l=1$ (dipole) modes with circular symbols, and the
  $l=2$ (quadrupole) modes with triangular symbols. Symbol sizes
  reflect the relative visibilities of the different modes, with a
  suitable correction included to reflect the impact of mixing on the
  mode inertia.  All model frequencies are plotted, irrespective of
  whether we were able to report a reliable observed frequency for
  them.}

\label{fig:osc}
\end{figure}



\begin{figure}
\centering
\includegraphics[width=165mm]{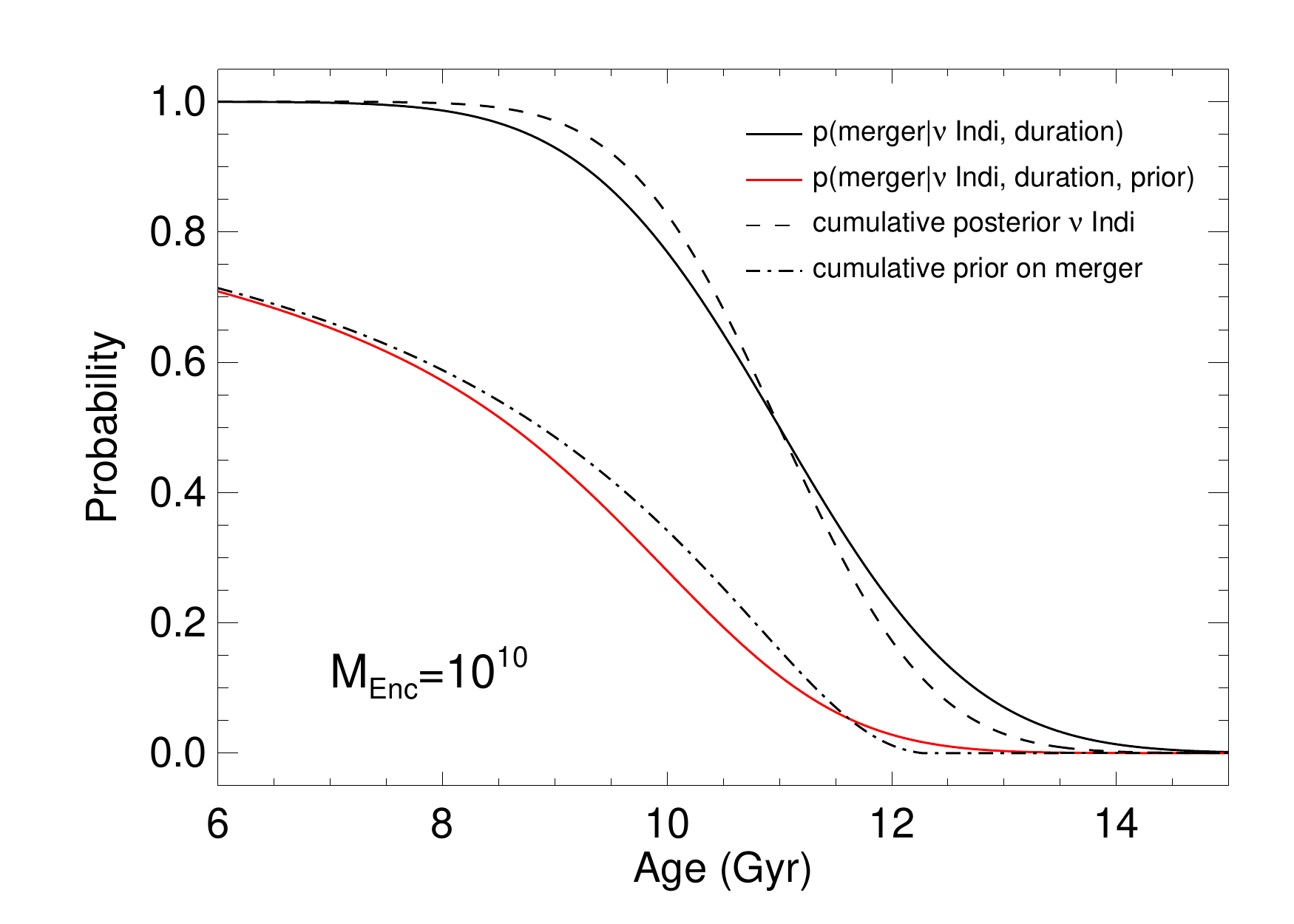}

\caption{\small Inference on the epoch of the \textit{Gaia}-Enceladus
  merger. The dashed black line shows the measured cumulative
  posterior on $\nu$\,Indi. The dot-dashed black line is the estimated
  cumulative prior probability for the merger assuming a virial mass
  of the \textit{Gaia}-Enceladus dark matter halo of $M_{\rm Enc} = 1
  \times 10^{10}\,\rm M_{\odot}$. The solid black line shows the
  cumulative probability for the merger, dependent on the estimated
  age of $\nu$\,Indi and the assumed 2-Gyr-wide merger duration; while
  the solid red line shows the cumulative probability for the merger
  also taking into account the merger prior (different in each panel,
  since this depends on $M_{\rm Enc}$).}

\label{fig:osc}
\end{figure}



\begin{figure}
\centering
\includegraphics[width=165mm]{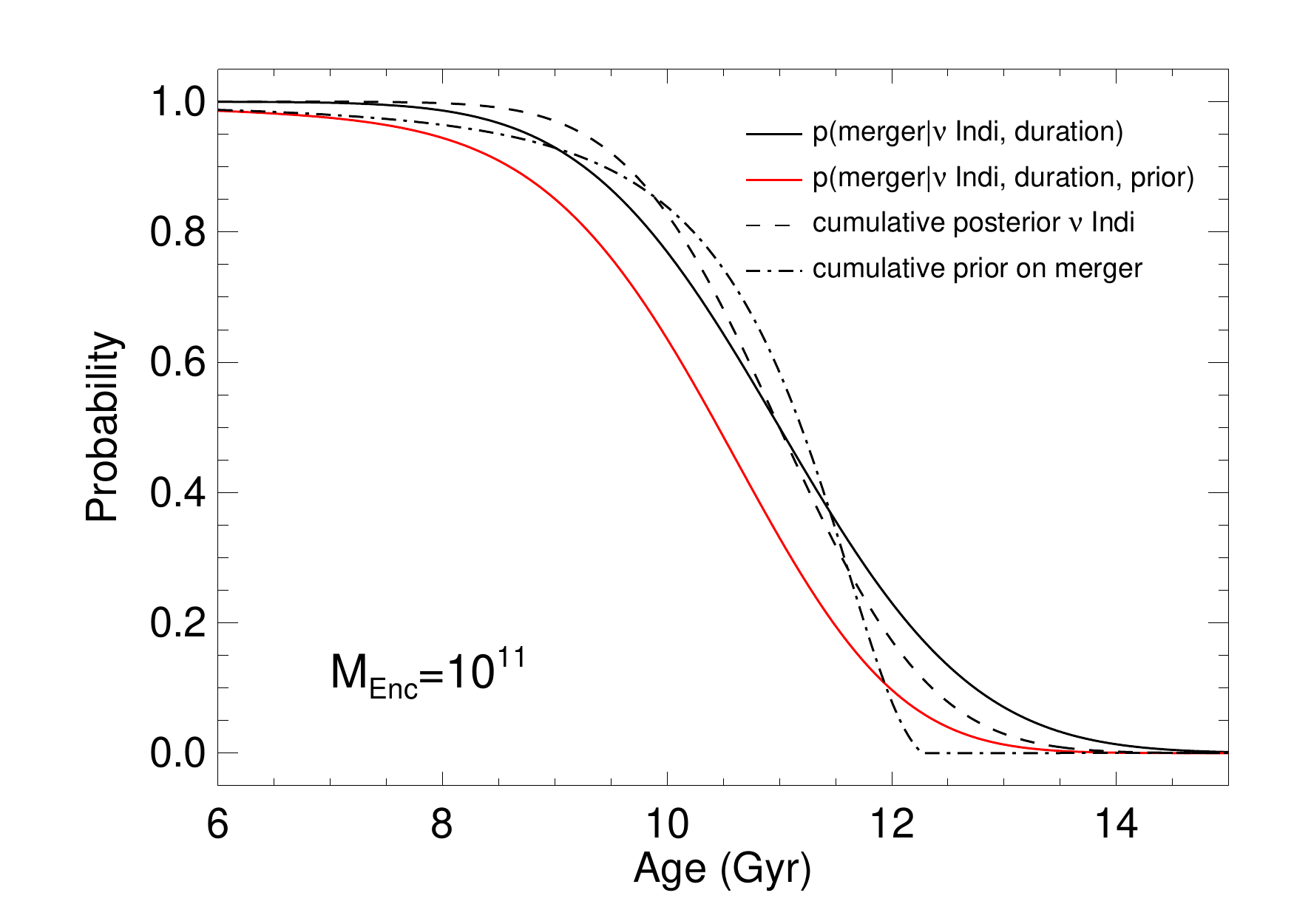}

\caption{\small As per Figure 6, but now assuming a
  virial mass of the \textit{Gaia}-Enceladus dark matter halo of $1
  \times 10^{11}\,\rm M_{\odot}$. [Note the measured cumulative
    posterior on $\nu$\,Indi (dashed black line) and the cumulative
    probability for the merger (dependent on the estimated age of
    $\nu$\,Indi and the assumed 2-Gyr-wide merger duration; black
    line) are the same as in Figure 6.]}

\label{fig:osc}
\end{figure}


\noindent \textbf{Data Availability}\\
Raw TESS data are available from the MAST portal at  {https://archive.stsci.edu/access-mast-data}. The TASOC lightcurve is available at  {https://tasoc.dk/}. The TESS lightcurve and power spectrum is also available on request from the corresponding author. The high-resolution spectroscopic data are available at  {http://archive.eso.org/wdb/wdb/adp/phase3{\_}spectral/form} (HARPS $\nu$\,Indi),  {https://www.blancocuaresma.com/s/benchmarkstars} (HARPS solar spectrum), and\\ 
{http://archive.eso.org/wdb/wdb/adp/phase3{\_}spectral/form} (FEROS). MARCS model atmospheres are available at  {http://marcs.astro.uu.se/}. APOGEE Data Release 14 may be accessed via\\  {https://www.sdss.org/dr14/}.\\

\noindent \textbf{Code Availability}\\
The adopted asteroseismic modelling results were provided by the BeSPP code, which is available on request from A.M.S. (aldos@ice.csic.es). NLTE corrections were estimated using the interactive online tool at http://nlte.mpia.de.  The computation of Kurucz models with ATLAS9 was performed using http://atmos.obspm.fr/index.php/documentation/7.  Publicly available codes used to model the data include IRAF (http://ast.noao.edu/data/software), MOOG\\ (https://www.as.utexas.edu/~chris/moog.html), the MCMC code \texttt{emcee} (https://github.com/dfm/emcee), the peak-bagging codes DIAMONDS (https://github.com/EnricoCorsaro/DIAMONDS) and TAMCMC-C (https://github.com/OthmanB/TAMCMC-C), the stellar evolution code MESA\\ (http://mesa.sourceforge.net/), and the stellar pulsation code GYRE\\ (https://bitbucket.org/rhdtownsend/gyre/wiki/Home). Other codes used in the analysis -- including frequency analysis tools -- are available on reasonable request via the corresponding author.\\

\end{document}